\documentclass[two-column, nofootinbib]{revtex4-1}

\usepackage{amsmath}
\usepackage{tabularx}
\usepackage{graphicx}
\usepackage{bigints}
\usepackage[usenames,dvipsnames,svgnames]{xcolor}
\usepackage{hyperref}

\hypersetup{dvips,dvipdfm,colorlinks=true,urlcolor=magenta,filecolor=magenta,linktoc=page,citecolor=red,linkcolor=blue,bookmarks=true}

\begin{document}

\title{Interacting Chaplygin gas revisited}
%\title{Can the Chaplygin gas and a barotropic fluid with a constant equation of state coexist?}
%\title{Can the Chaplygin gas befriend a barotropic fluid with a constant equation of state?}

%\author{}}
%\affiliation{}

\author{Subhajit Saha\footnote {Electronic Address: \texttt{\color{blue} subhajit1729@gmail.com}}}
%\affiliation{Department of Physical Sciences, \\ Indian Institute of Science Education and Research Kolkata, Mohanpur 741246, West Bengal, India} 
\affiliation{Department of Mathematics, \\ Panihati Mahavidyalaya, \\ Sodepur 700110, West Bengal, India.} 
\author{Saumya Ghosh\footnote {Electronic Address: \texttt{\color{blue} sgsgsaumya@gmail.com}}}
\author{Sunandan Gangopadhyay\footnote {Electronic Address: \texttt{\color{blue} sunandan.gangopadhyay@gmail.com}}}
\affiliation{Department of Physical Sciences, \\ Indian Institute of Science Education and Research Kolkata, Mohanpur 741246, West Bengal, India.} 

%%%%%%%%%%%%%%%%%%%%%%%%%%%%%%%%%%%%%%%%%%%%%%%%%%%%%%%%%%%%%%%%%%%%%%%%%%%%%%%%%%%%%%%%%%%%%%%%%%%%%%%%%%%%%%%%%%%%%%%%%%%%

\begin{abstract}

\begin{center}
(Dated: The $23^{\text{rd}}$ June, $2017$)
\end{center}

The implications of considering interaction between Chaplygin gas and a barotropic fluid with constant equation of state have been explored. The unique feature of this work is that assuming an interaction $Q \propto H\rho_d$, analytic expressions for the energy density and pressure have been derived in terms of the Hypergeometric $_2\text{F}_1$ function. It is worthwhile to mention that an interacting Chaplygin gas model was considered in 2006 by Zhang and Zhu, nevertheless, analytic solutions for the continuity equations could not be determined assuming an interaction proportional to $H$ times the sum of the energy densities of Chaplygin gas and dust. Our model can successfully explain the transition from the early decelerating phase to the present phase of cosmic acceleration. Arbitrary choice of the free parameters of our model through trial and error show at recent observational data strongly favors $w_m=0$ and $w_m=-\frac{1}{3}$ over the $w_m=\frac{1}{3}$ case. Interestingly, the present model also incorporates the transition of dark energy into the phantom domain, however, future deceleration is forbidden. \\\\ 
Keywords: Chaplygin gas; Barotropic fluid; Interaction; Analytic solution\\\\
PACS Numbers: 98.80.-k, 95.35.+d, 95.36.+x \\\\

\end{abstract}

\maketitle

%%%%%%%%%%%%%%%%%%%%%%%%%%%%%%%%%%%%%%%%%%%%%%%%%%%%%%%%%%%%%%%%%%%%%%%%%%%%%%%%%%%%%%%%%%%%%%%%%%%%%%%%%%%%%%%%%%%%%%%%%%%%

%%%%%%%%%%%%%%%%%%%%%%%%%%%%%%%%%%%%%%%%%%%%%%%%%%%%%%%%%%%%%%%%%%%%%%%%%%%%%%%%%%%%%%%%%%%%%%%%%%%%%%%%%%%%%%%%%%%%%%%%%%%%

\section*{1. Introduction}

Chaplygin gas (CG) is a class of dark energy (DE) models (for a review on DE, see \cite{Copeland1,Amendola0}) which mimics the observed late-time acceleration of the Universe as was reported in \cite{Riess1,Perlmutter1,Bachall1}. Kamenshchik, Moschella, and Pasquier \cite{Kamenshchik1} were the first to consider CG in the context of Cosmology. Such a fluid is assumed to have the following equation of state (EoS)\footnote{An ``anti-Chaplygin" EoS (Eq. (\ref{cgeos}) with a negative $A$) can be found in the description of wiggly strings \cite{Carter1,Vilenkin1}.}:
\begin{equation} \label{cgeos}
p=-\frac{A}{\rho},
\end{equation}
where $p$ and $\rho$ are, respectively, pressure and energy density in a comoving reference frame, and $A$ is a positive constant. It is named after Sergey Chaplygin\footnote{Sergey Chaplygin was a Russian mathematician, physicist, and engineer who is very likely the only scientist who has a lunar crater, a city, and a cosmological model named after him.} who introduced the above EoS \cite{Chaplygin1} as a suitable mathematical approximation for calculating the lifting force on a plane wing in aerodynamics. CG can also be considered as a special case of a tachyon with a constant potential \cite{Frolov1,Copeland1}.

The convenience of CG is due to the fact that the corresponding Euler equations have a very large group of symmetry, which implies their integrability. The relevant symmetry group has been recently described in modern terms \cite{Bazeia1}. CG has a remarkable connection with string theory since it can be obtained from the Nambu-Goto action for $D$-branes moving in a $(D+2)$-dimensional spacetime in the light-cone parametrization \cite{Bordemann1}. It can also be derived for the moving brane via the Born-Infeld Lagrangian \cite{Bento1}. Further, CG is also known to admit a supersymmetric generalization \cite{Hoppe1,Jackiw1}. Certain effects in deformable solids, of stripe states in the context of the quantum Hall effect and of other phenomena, can be explained by the negative pressure arising from the CG EoS \cite{Stanyukovich1}. It is also useful when studying the stabilization of branes in black hole bulks \cite{Kamenshchik2,Randall1,Banados1}. A microscopic description of CG suggests that it can be regarded (phenomenologically) as the effect of the immersion of our four-dimensional world into some multidimensional bulk \cite{Bilic1,Sundrum1}. Another intriguing property of CG is that it gives positive and bounded square of sound velocity $c_{s}^{2}=\frac{A}{\rho ^2}$, which is a non-trivial fact for fluids with negative pressure \cite{Kamenshchik1,Gorini1}. This sound velocity is negligible at early times and approaches the speed of light in the late-time limit (for a detailed discussion on this topic, see \cite{Zimdahl1}).

The advantages \cite{Kamenshchik1,Gorini1} of the Chaplygin class of cosmological models is three-fold. Firstly, they describe a smooth transition from the decelerating phase of the Universe to the present phase of cosmic acceleration and such a transition is achieved with only one fluid. Secondly, a unified macroscopic phenomenological description of DE and dark matter (DM) can be given by such a class of models. Finally, they represent the simplest deformations of the concordance $\Lambda$CDM model. In spite of these remarkable advantages, CG is in disagreement with the observational data obtained from CMB anisotropies \cite{Bento1,Bilic2,Amendola1} which is attributed to the fact that the Jeans instability of perturbations in CG models behaves similarly to cold dark matter (CDM) fluctuations in the dust-dominant stage but disappears in the acceleration stage. A strong integrated Sachs-Wolfe (ISW) effect arises due to the dual effect of the suppression of perturbations and the presence of a non-zero Jeans length \cite{Copeland1,Zimdahl1}. To remedy the situation, the generalized Chaplygin gas (GCG) model was proposed \cite{Bento1} which has an EoS given by $p=-\frac{B}{\rho ^\alpha}$ with $B>0$ and $0<\alpha <1$, nevertheless, the parameter $\alpha$ is severely constrained, i.e., $0<\alpha <0.2$ at the 95\% confidence level \cite{Bilic1}. Later, modified Chaplygin gas (MCG) model was proposed \cite{Benaoum1} which is an extension of the GCG model with EoS as $p=C\rho-\frac{D}{\rho ^n}$, $C$ \& $D$ are positive constants and $n \geq 1$. Various other modifications of CG have appeared in the literature such as variable CG \cite{Guo1}, holographic and interacting holographic CG \cite{Setare1,Setare2}, viscous CG models (first proposed in \cite{Zhai1}) amongst others, however, each one of them comes with both merits and demerits as far as Cosmology is concerned. CG models have also been considered in modified as well as higher-dimensional gravity theories. 

In the present work, a flat Friedmann-Lemaitre-Robertson-Walker (FLRW) universe has been considered and assumed to be filled with two fluids --- CG and a barotropic fluid with EoS $p=w\rho$, $w$ constant. Our investigation has been primarily focussed on the dynamics of the coexistence of the fluids %(i) without interaction, and (ii) 
in the presence of an interaction term proportional to the Hubble parameter times the DE density. %(in this case, the matter field has been assumed to be in dust form for simplicity). 
This class of interaction terms generally appears in the interacting holographic dark energy (HDE) model \cite{Kim1,Wang1}. In the absence of interaction, there exists no scaling solutions owing to the fact that the EoS of CG decreases with scale factor while the dark matter (DM) EoS remains constant. It is worthwhile to mention that Zhang and Zhu \cite{Zhang1} have earlier considered a kind of interacting Chaplygin gas model in which the Chaplygin gas plays the role of DE and interacts with CDM particles via an interaction term of the form $\Gamma=3cH(\rho_d+\rho_m)$, where $\rho_d$ and $\rho_m$ respectively denote the energy densities due to DE and CDM, $H$ is the Hubble parameter, and $c$ is the coupling constant. They found a stable scaling solution at late times with the Universe evolving into a phase of steady state. Moreover, their effective EoS could also cross the phantom barrier. However, their form of interaction failed to produce an analytic solution which is necessary to obtain in order to have a clear and a nice picture of the cosmological model concerned.

Our paper is organized as follows --- Section 2 describes the basic equations that govern a flat FLRW universe filled with CG and a fluid with EoS $p=w\rho$, $w$ constant. %The dynamics associated with such a cosmological scenario in the absence of an interaction term has been presented in Section 3. 
Section 3 is concerned with the cosmological implications of considering interaction between the two sectors. Finally, a short discussion and scope of future work appears in Section 4.

\section*{2. Basic equations of our model}

As stated, we consider a flat FLRW universe governed by the metric
\begin{equation} \label{flrw}
ds^2=-dt^2+a^2(t)\left[dr^2+r^2\left(d\theta ^2 +\text{sin}^2\theta d\phi ^2\right)\right],
\end{equation}
where $a(t)$ is the scale factor of the Universe. Assuming a perfect fluid having energy-momentum tensor given by ($u_\mu$ is the $4$-velocity of the fluid)
\begin{equation}
T_{\mu \nu}=(\rho +p)u_{\mu}u_{\nu}+pg_{\mu \nu},
\end{equation}
the Friedmann and the acceleration equations can be obtained as
\begin{eqnarray}
3H^2 &=& \rho, \label{fe} \\
2\dot{H} &=& -(\rho +p) \label{ae}
\end{eqnarray}
respectively, where $H$ is the Hubble parameter defined by $H=\frac{\dot{a}(t)}{a(t)}$, $\rho$ is the total energy density of the Universe and $p$ is the pressure term. We have also assumed that $8\pi G=c=1$, without any loss of generality. Using the above equations, one can obtain the energy-momentum conservation equation 
\begin{equation} \label{ce}
\dot{\rho}+3H(\rho +p)=0.
\end{equation}

Since we shall be working with a two-fluid system, the total energy density $\rho$ and the total pressure $p$ can be written as
\begin{eqnarray}
\rho &=& \rho_m+\rho_d \label{rho} \\
p &=& p_m+p_d, \label{p}
\end{eqnarray}
where $\rho_d$ and $p_d$ represent, respectively, the energy density and pressure due to DE which is considered to be CG (EoS: $p_d=-\frac{A}{\rho _d}$, $A$ is a positive constant), and the corresponding quantities with suffix $m$ are due to the matter field which we shall assume to be a barotropic fluid with EoS given by $p_m=w_m\rho_m$, $w_m \geq -\frac{1}{3}$ is a constant. The lower bound on $w_m$ assures that the barotropic fluid does not violate the strong energy condition.

\section*{3. Cosmological dynamics for $\text{CG}+\text{BAROTROPIC FLUID}$ with interaction $Q \propto H\rho_d$} 

We shall now study the implications of considering interaction between matter (barotropic fluid with constant EoS) and DE (CG) sectors. It has already been mentioned earlier that an interacting CG model was considered in the literature \cite{Zhang1}, nevertheless, analytic solutions for the continuity equations could not be determined assuming an interaction of the form $\Gamma=3cH(\rho_d+\rho_m)$. The authors \cite{Zhang1} performed a phase-space analysis and obtained a stable scaling solution at late times with the Universe evolving into a phase of steady state. Since there is no microphysical hint on the nature of interaction between Chaplygin gas and matter, we are bound to consider a phenomenological form of the interaction term. In what follows, it can be seen that the non-conservation equations for CG and the barotropic fluid arising due to an interaction of the form $Q=3b^2H\rho_d$ ($b^2$ is the coupling parameter) produces analytic expressions for $\rho$ and $p$. 

\begin{figure}[h]
\begin{center}
\begin{minipage}{0.315\textwidth}
\includegraphics[width=1.0\linewidth]{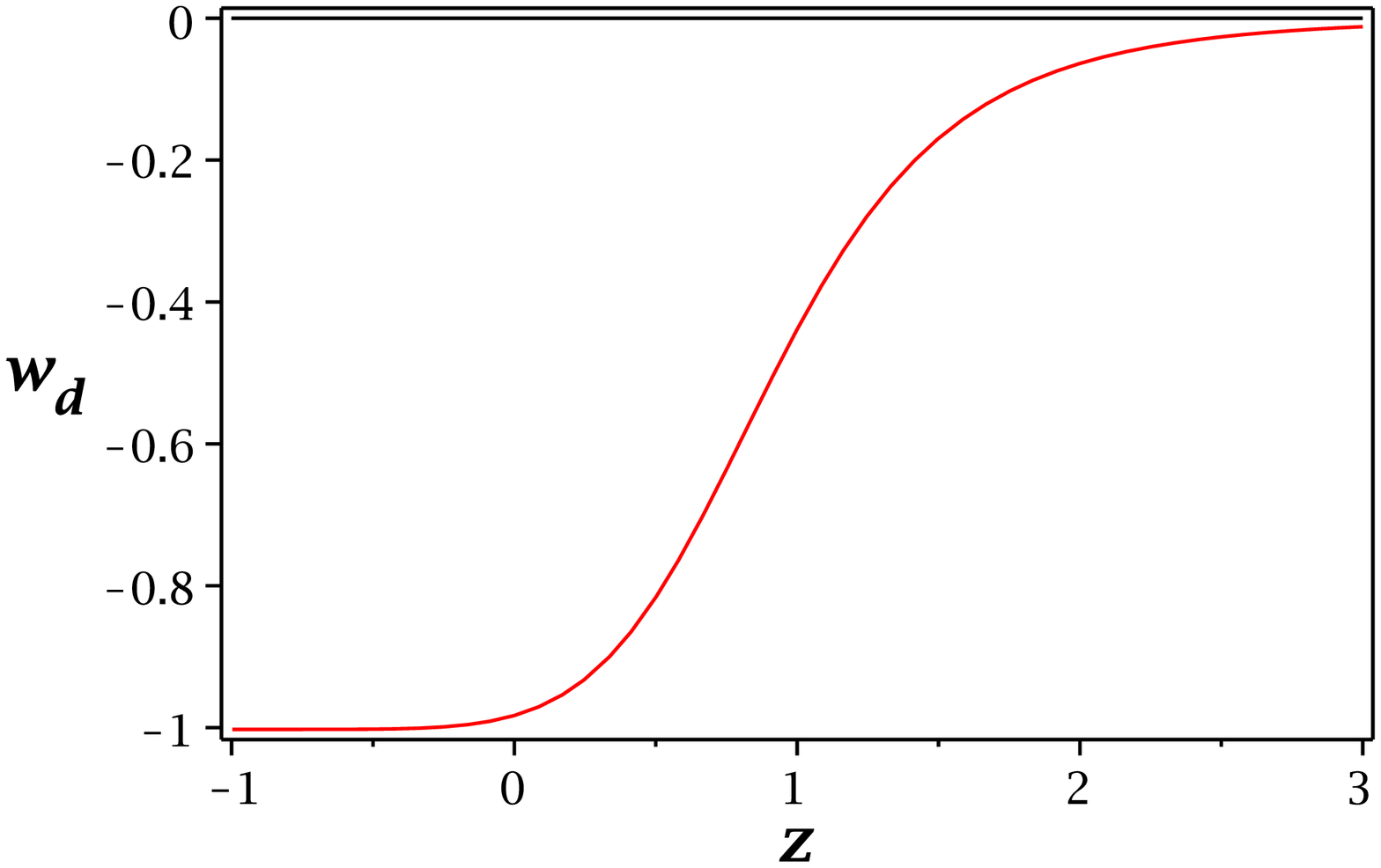}
\end{minipage}
\hspace*{0.15cm}
\begin{minipage}{0.315\textwidth}
\includegraphics[width=1.0\linewidth]{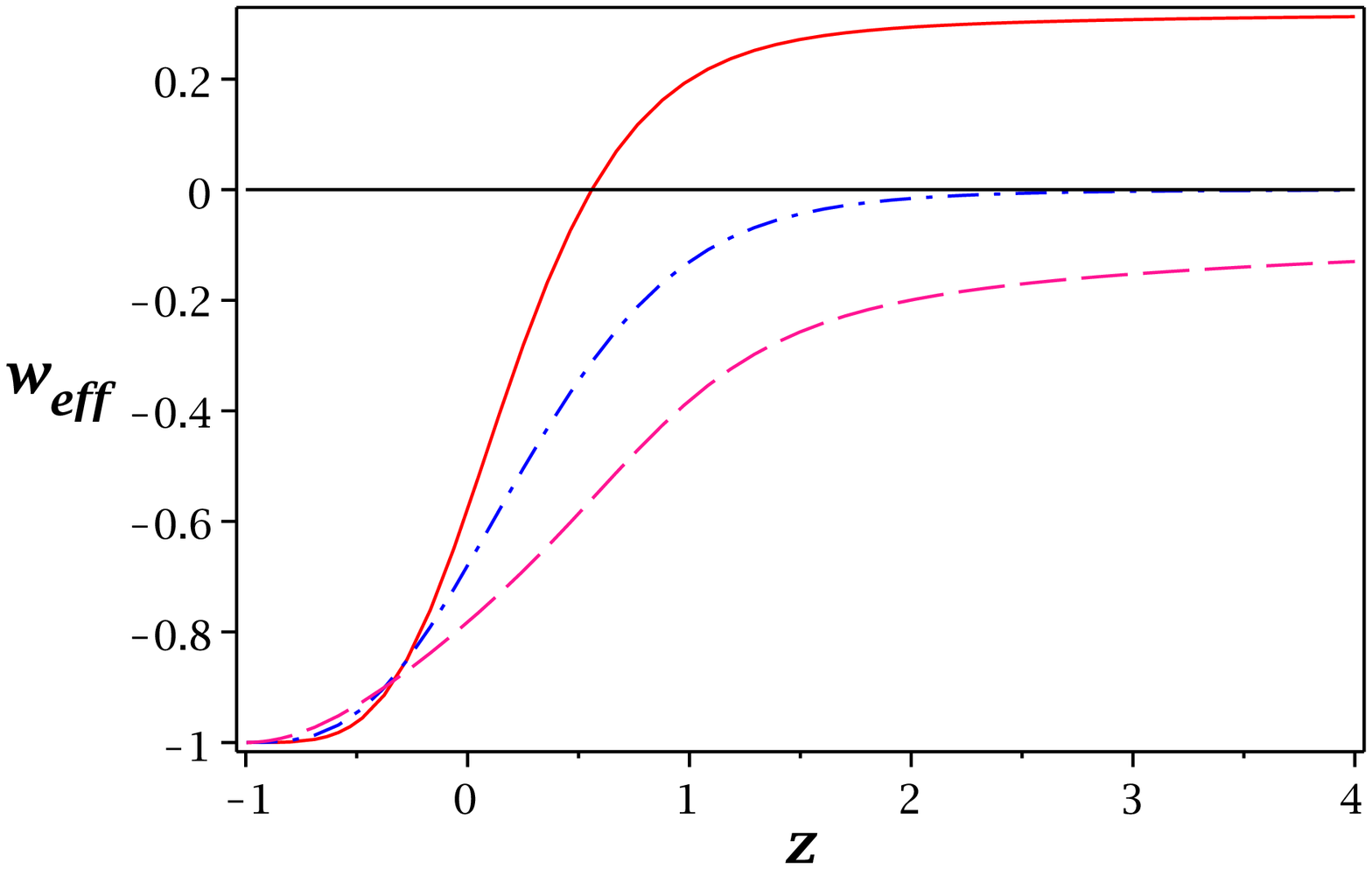}
\end{minipage}
\hspace*{0.15cm}
\begin{minipage}{0.315\textwidth}
\includegraphics[width=1.0\linewidth]{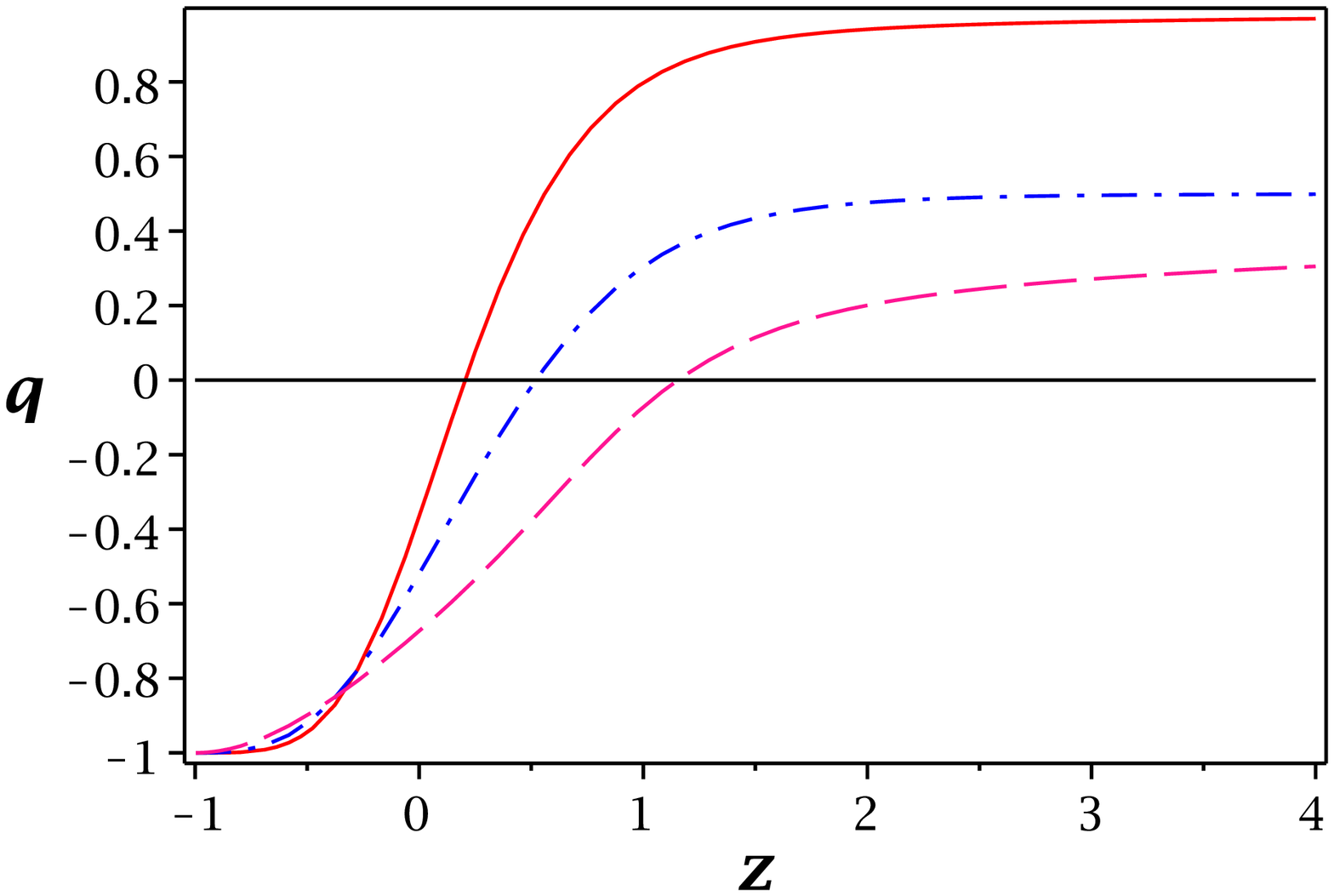}
\end{minipage}
\caption{The variations of the DE density $w_d$, the effective EoS $w_{eff}$, and the deceleration parameter $q$ against the redshift $z$. The solid, dashdot, and dashed curves in the middle and the right panels correspond to $w_m=\frac{1}{3}$, $w_m=0$, and $w_m=-\frac{1}{3}$ respectively.}
\label{Fig:1}
\end{center}
\end{figure}

Interaction between DE and DM has some important consequences such as in alleviating the coincidence problem \cite{Amendola1,Chimento1,Olivares1,Mangano1,Farrar1,Pavon1}, among others. The coincidence problem can be solved, or atleast alleviated if DE decays into DM \cite{Zhang1}, thus diminishing the difference between the densities of the two components through the evolution of the Universe. Therefore, if CG is assumed to decay into matter, then under the said form of interaction, the non-conservation equations for the two sectors can be written as
\begin{eqnarray}
\dot{\rho}_d+3H(p_d+{\rho_d}) &=& -Q = -3b^2H\rho_d, \label{cedemod} \\
\dot{\rho}_m+3(1+w_m)H\rho_m &=& +Q = +3b^2H\rho_d. \label{cemmod}
\end{eqnarray}
Plugging in the EoS of CG in Eq. (\ref{cedemod}), and evaluating
\begin{equation}
\bigint \frac{\text{d}\rho_d}{\rho_d \left(1+b^2-\frac{A}{\rho_{d}^{2}}\right)}=\text{ln}\left(\frac{B'}{a^3}\right), \nonumber
\end{equation} 
we obtain $\rho_d$ as
\begin{equation}
\rho_d = \frac{1}{\sqrt{1+b^2}}\left[A+\left(\frac{B'}{a^6}\right)^{1+b^2}\right]^{\frac{1}{2}}, \label{rhodint}
\end{equation}
where $B'$ is the constant of integration. Note that when there is no interaction, i.e., $b^2=0$, we get back the CG density obtained in \cite{Kamenshchik1}. Now, putting the above expression for $\rho_d$ in Eq. (\ref{cemmod}) and multiplying both sides of the equation by $a^{3(1+w_m)}$, the matter density $\rho _m$ can be evaluated as\footnote{Mathematica software was used to evaluate the integral obtained in Eq. (\ref{cemmod}).}
\begin{eqnarray}
\rho_m &=& \frac{1}{a^{3(1+w_m)}}\Biggl[C'+\frac{b^2}{(b^2-w_m)(1+w_m)\sqrt{(1+b^2)\left[A+\left(\frac{B'}{a^6}\right)^{1+b^2}\right]}}\Bigg\lbrace a^{3(1+w_m)}\Biggl(-(1+w_m) \left\lbrace A+\left(\frac{B'}{a^6}\right)^{1+b^2} \right\rbrace \nonumber \\
&+& \sqrt{A}(1+b^2)\sqrt{A+\left(\frac{B'}{a^6}\right)^{1+b^2}}\times {_2\text{F}_{1}}\left[\frac{1}{2},-\frac{1+w_m}{2(1+b^2)},1-\frac{1+w_m}{2(1+b^2)},-\frac{1}{A}{\left(\frac{B'}{a^6}\right)}^{1+b^2}\right]\Biggr)\Bigg\rbrace \Biggr] \label{rhomint}
\end{eqnarray}
from the integral
\begin{equation}
a^3 \rho_m=\frac{3b^2}{\sqrt{1+b^2}}\bigint a^2\sqrt{A+\left(\frac{B}{a^6}\right)^{1+b^2}}\text{d}a+C', \nonumber
\end{equation}
with the constant of integration as $C'$. Few comments\footnote{The authors are grateful to the anonymous reviewer for raising this very important issue.} regarding the free parameters of our model are in order. Note that for a prescribed matter EoS $w_m$, our model consists of four free parameters --- the CG parameter $A$, the coupling parameter $b^2$, and the two constants of integration $B'$ and $C'$. If one restricts to the spatially flat case, then $\Lambda$CDM has only one free parameter ($\Omega _{m0}$), while the most discussed dynamical DE model, $\phi$CDM has two free parameters ($\Omega _{m0}$ and $\alpha$) \cite{Peebles1,Ratra1}. In the latter case, one has the attractor solution, so there is no dependence on initial conditions. In our interacting CG model, the two constants arising due to integration can be fixed so that we shall also be left with only two free parameters, $A$ and $b^2$. However, due to a high degree of nonlinearity in the expressions, it is quite difficult to identify the relations of the parameters with those occurring in the more well-known DE models. Now, the explicit expressions for the total energy density $\rho$ and the total pressure $p$ can be written as
\begin{eqnarray}
\rho &=& \frac{1}{a^{3(1+w_m)}}\Biggl[C'+\frac{b^2}{(b^2-w_m)(1+w_m)\sqrt{(1+b^2)\left[A+\left(\frac{B'}{a^6}\right)^{1+b^2}\right]}}\Bigg\lbrace a^{3(1+w_m)}\Biggl(-(1+w_m) \left\lbrace A+\left(\frac{B'}{a^6}\right)^{1+b^2} \right\rbrace \nonumber \\
&+& \sqrt{A}(1+b^2)\sqrt{A+\left(\frac{B'}{a^6}\right)^{1+b^2}}\times {_2\text{F}_{1}}\left[\frac{1}{2},-\frac{1+w_m}{2(1+b^2)},1-\frac{1+w_m}{2(1+b^2)},-\frac{1}{A}{\left(\frac{B'}{a^6}\right)}^{1+b^2}\right]\Biggr)\Bigg\rbrace \Biggr] \nonumber \\
&+& \frac{1}{\sqrt{1+b^2}}\left[A+\left(\frac{B'}{a^6}\right)^{1+b^2 }\right]^{\frac{1}{2}}, \label{rhoint}
\end{eqnarray}
\begin{eqnarray}
p &=& \frac{w_m}{a^{3(1+w_m)}}\Biggl[C'+\frac{b^2}{(b^2-w_m)(1+w_m)\sqrt{(1+b^2)\left[A+\left(\frac{B'}{a^6}\right)^{1+b^2}\right]}}\Bigg\lbrace a^{3(1+w_m)}\Biggl(-(1+w_m) \left\lbrace A+\left(\frac{B'}{a^6}\right)^{1+b^2} \right\rbrace \nonumber \\
&+& \sqrt{A}(1+b^2)\sqrt{A+\left(\frac{B'}{a^6}\right)^{1+b^2}}\times {_2\text{F}_{1}}\left[\frac{1}{2},-\frac{1+w_m}{2(1+b^2)},1-\frac{1+w_m}{2(1+b^2)},-\frac{1}{A}{\left(\frac{B'}{a^6}\right)}^{1+b^2}\right]\Biggr)\Bigg\rbrace \Biggr] \nonumber \\
&-& A\sqrt{(1+b^2)} \left[A+\left(\frac{B'}{a^6}\right)^{1+b^2}\right]^{-\frac{1}{2}} \label{pint}
\end{eqnarray}
respectively, where $_2\text{F}_1[y_1,y_2,y_3,x]$ is known as the Gauss's hypergeometric function \cite{http}. The effective EoS parameter $w_{eff}=\frac{p}{\rho}$ and the deceleration parameter $q=\frac{3}{2}\left(1+\frac{p}{\rho}\right)-1$ for this interacting scenario can also be easily constructed using Eqs. (\ref{rhoint}) and (\ref{pint}). We do not write them explicitly in order to avoid unnecessary expansion of the manuscript. Since the above expressions are quite complicated, it is very difficult to analyze the present model analytically. Instead, the variations of the relevant parameters, namely, $w_d$, $w_{eff}$, and $q$ against the redshift $z$ have been presented in Figure \ref{Fig:1}. In doing so, we have assumed the following values for different free parameters: $A=5$, $b=0.05$, $B'=0.1$, and $C'=1$. These choices lead to a CG EoS of $-0.983$ at the present epoch which is nearly consistent with recent observations \cite{Ade1}. The values of the other parameters, particularly, $q$, $\Omega_d$, and $\Omega_m$ at the present epoch (for the three chosen values of $w_m$) have been presented in Table \ref{Tab:1}. Among the three cases, the values corresponding to $w_m=0$ and $w_m=-\frac{1}{3}$ show a better consistency with the latest observational data \cite{Ade1,Santos1,Campo1,Akarsu1,Nair1,Mamon1} compared to the $w_m=\frac{1}{3}$ case. We have also calculated the redshift of transition from deceleration to acceleration ($z_{da}$) for $w_m=\frac{1}{3}$, $w_m=0$, and $w_m=-\frac{1}{3}$ to be $\approx 0.20$, $\approx 0.52$, and $\approx 1.14$ respectively. The dust case shows a better agreement with previous analyses \cite{Farooq1,Farooq2} as compared to the other two cases. A chi-square analysis could have been more fruitful in this situation but due to a large number of free parameters, the software packages were not able to produce interesting results.

%The present values of $w_d$, $w_{eff}$, and $q$ obtained with the assumed values of the free parameters are as follows: $w_d=-0.975$; $w_{eff}=-0.522$ ($w_m=\frac{1}{3}$), $w_{eff}=-0.637$ ($w_m=0$), $w_{eff}=-0.752$ ($w_m=-\frac{1}{3}$); $q=-0.283$ ($w_m=\frac{1}{3}$), $q=-0.456$ ($w_m=0$), $q=-0.628$ ($w_m=-\frac{1}{3}$). 

\begin{table}[h] 
\caption{Present values of the cosmological parameters with $A=5$, $b=0.05$, $B'=0.1$, and $C'=1$}
\begin{center}
\begin{tabular}{p{2cm}p{1.75cm}p{1.75cm}p{1.75cm}}
\hline \hline Parameter & $w_m=\frac{1}{3}$ & $w_m=0$ & $w_m=-\frac{1}{3}$ \\
%\hline \\ $w_{eff}$ & $-0.577$ & $-0.680$ & $-0.782$ \\\\
\hline \\ $q$ & $-0.366$ & $-0.520$ & $-0.673$ \\\\
 	   $\Omega_d$ & $~~0.692$ & $~~0.692$ & $~~0.691$ \\\\
 	   $\Omega_m$ & $~~0.308$ & $~~0.308$ & $~~0.309$ \\\\
\hline \hline
\end{tabular}
\end{center}
\label{Tab:1}
\end{table}

Unlike the non-interacting CG models, this interacting scenario incorporates the transition of DE into the phantom regime. Since
\begin{eqnarray}
w_d &=& -\frac{A}{\rho _{d}^{2}} \nonumber \\
&=& -\frac{A(1+b^2)}{A+\left(\frac{B'}{a^6}\right)^{1+b^2 }},
\end{eqnarray}
we observe that as the scale factor becomes very large, the second term inside the denominator can be ignored and $w_d$ reduces to $w_d=-1-b^2$ which always has a value less than the phantom barrier of $-1$, for a non-vanishing coupling parameter $b^2$.\\

\begin{figure}[h]
\begin{center}
\begin{minipage}{0.3\textwidth}
\includegraphics[width=1.0\linewidth]{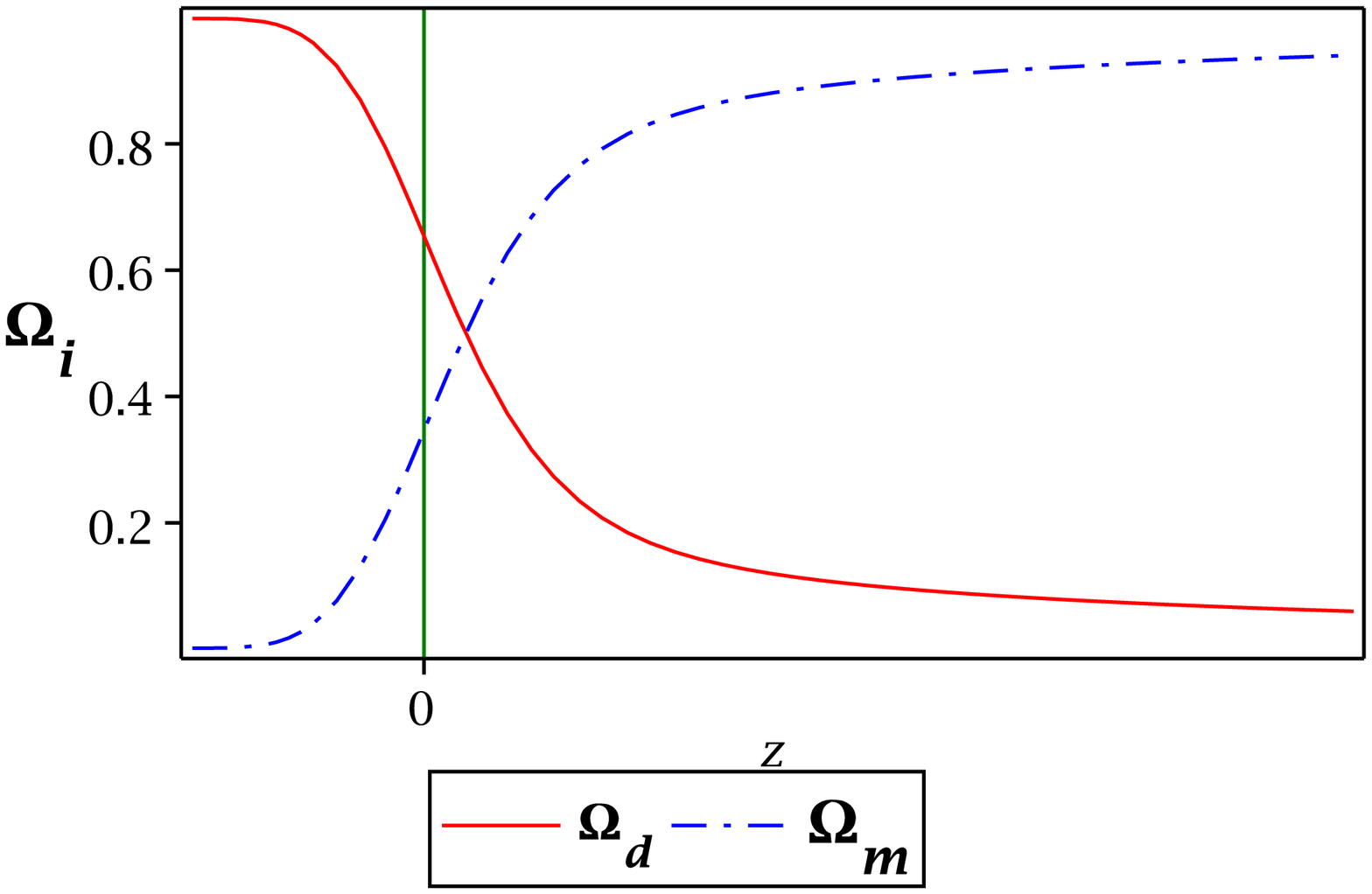}
\end{minipage}
\hspace*{0.25cm}
\begin{minipage}{0.3\textwidth}
\includegraphics[width=1.0\linewidth]{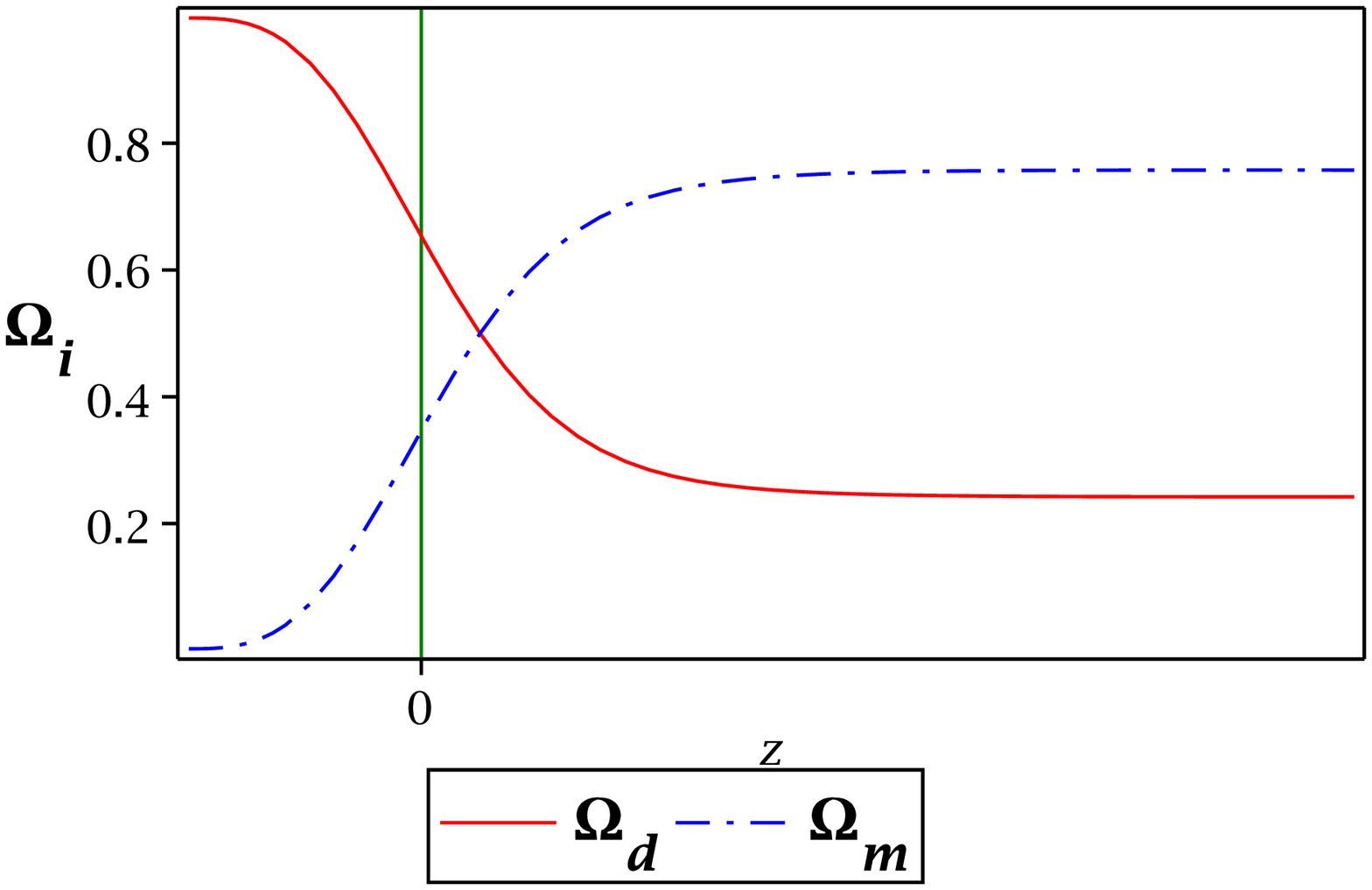}
\end{minipage}
\hspace*{0.25cm}
\begin{minipage}{0.3\textwidth}
\includegraphics[width=1.0\linewidth]{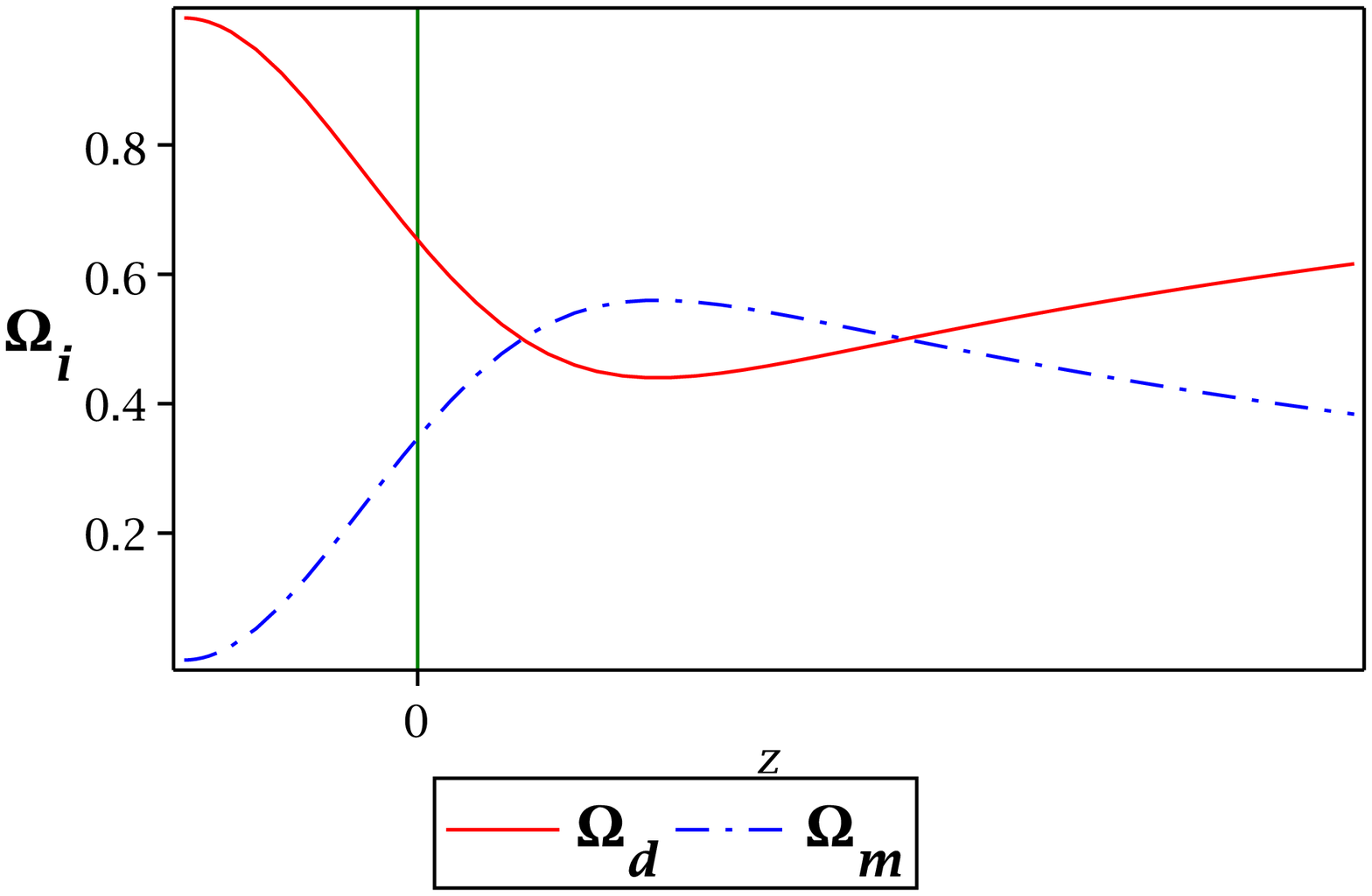}
\end{minipage}
\caption{The variations of the fractional energy densities of DE ($\Omega_d$) and matter ($\Omega_m$) against the redshift $z$. The left, middle, and right panels correspond to $w_m=\frac{1}{3}$, $w_m=0$, and $w_m=-\frac{1}{3}$ respectively.}
\label{Fig:2}
\end{center}
\end{figure}

Using the same set of values for the free parameters, we have plotted the variations of the fractional\footnote{The fractional energy densities of $\Omega_d$ and $\Omega_m$ are given by $\Omega_d=\frac{\rho_d}{\rho}$ and $\Omega_m=\frac{\rho_m}{\rho}$ respectively.} energy densities of DE ($\Omega_d$) and matter ($\Omega_m$) and presented them in Figure \ref{Fig:2}. All the three panels show that CG has started dominating over the matter sector in recent past which has led to the observed late time acceleration of the Universe. It is also evident that the energy density of DE will steadily increase with the evolution of the Universe and will lead to complete evaporation of matter in some future epoch. However, there is some peculiarity in the $w_m=-\frac{1}{3}$ case, or in a more general sense, the negative $w_m$ case. The energy densities $\Omega_d$ and $\Omega_m$ form a "knot" in some past redshift interval (the position of the knot depends upon the values of the free parameters chosen) which implies that the CG was the dominant force in the early Universe. However, it is not correct to speculate that it could explain the inflationary stage because in that case, $q$ should have shown two transitions, which is of course not evident from the variation of $q$ in Figure \ref{Fig:1}. Therefore, we can conclude that in the presence of an interaction of the form $Q=3b^2H\rho_d$, CG and the barotropic fluid with EoS $p_m=w_m\rho_m$ $\left(w_m \geq -\frac{1}{3}\right)$ can be considered as suitable candidates for DE and matter respectively. In other words, this is a suitable model to explain the medieval deceleration phase as well as the late-time acceleration phase of the Universe. Nevertheless, this type of interaction does not support future deceleration as has been reported in certain particle creation and backreaction models \cite{Bose1,Chakraborty1}. %Lastly, the coupling parameter $b^2$ should also be very small, otherwise, $\Omega _d$ becomes negative at a high redshift. 

%\begin{figure}[h]
%\begin{center}
%\includegraphics[width=0.4\linewidth]{M2_q+weff.eps}
%\caption{}
%\label{Fig:}
%\end{center}
%\end{figure}

%\begin{figure}[h]
%\begin{center}
%\includegraphics[width=0.4\linewidth]{M2_wd.eps}
%\caption{}
%\label{Fig:}
%\end{center}
%\end{figure}

%\begin{figure}[h]
%\begin{center}
%\includegraphics[width=0.4\linewidth]{M2_omd-omm.eps}
%\caption{}
%\label{Fig:}
%\end{center}
%\end{figure}

%\begin{figure}[h]
%\begin{center}
%\includegraphics[width=0.4\linewidth]{M2_sff.eps}
%\caption{}
%\label{Fig:}
%\end{center}
%\end{figure}

%\begin{figure}[h]
%\begin{center}
%\includegraphics[width=0.4\linewidth]{M1-M2_q.eps}
%\caption{}
%\label{Fig:}
%\end{center}
%\end{figure}

\section*{4. Short Discussion and Scope of Future work}

The consequences of considering a particular form of interaction (proportional to the Hubble parameter times the DE density) between CG and a barotropic fluid having constant EoS has been discussed. We have obtained analytic solutions (in terms of the Hypergeometric $_2\text{F}_1$ function) for the total energy density and the total pressure in the presence of such an interaction term. Interacting CG models have occurred in the literature (\cite{Zhang1} for instance), however, analytic solutions for the continuity equations could not be found assuming an interaction term proportional to Hubble parameter times the total energy density. Also, all of these models consider only dust as the matter source. Arbitrary choice of the free parameters of our model through trial and error show that recent observational data strongly favors $w_m=0$ and $w_m=-\frac{1}{3}$ over the $w_m=\frac{1}{3}$ case. The present model also shows the transition of DE into the phantom era in the future which is a property shared by the model of Zhang and Zhu \cite{Zhang1}. %It is also evident that the interaction term assumed by us is inconsistent with a negative EoS parameter for the barotropic fluid. 
However, future deceleration is not supported by our model. We reiterate that the merit of our work is in the analytic solution obtained with our assumption of the form of interaction which will provide a deeper picture of the model. This can be achieved by constraining the model parameters with the help of sophisticated data analysis softwares which can be the basis of a future work. %Future work in this direction may include constraining the model parameters with recent observational data sets. 
It would also be interesting to investigate the dynamics of an universe filled with CG and a barotropic fluid and interacting via various other forms of the interaction term which occur in the literature.

\begin{acknowledgments}
Subhajit Saha was partially supported by SERB, Govt. of India under National Post-doctoral Fellowship Scheme [File No. PDF/2015/000906]. Sunandan Gangopadhyay acknowledges the support by DST SERB under Start Up Research Grant (Young Scientist), File No. YSS/2014/000180.
\end{acknowledgments}

%%%%%%%%%%%%%%%%%%%%%%%%%%%%%%%%%%%%%%%%%%%%%%%%%%%%%%%%%%%%%%%%%%%%%%%%%%%%%%%%%%%%%%%%%%%%%%%%%%%%%%%%%%%%%%%%%%%%%%%%%%%%

\end{document}